# Focused Space Weather Strategy for Securing Earth, and Human Exploration of the Moon and Mars

A. Posner, N. Arge, K. Cho, B. Heber, F. Effenberger, T. Y. Chen, S. Krucker, P. Kühl,
O. Malandraki, Y.-D. Park, A. Pulkkinen, N. Raouafi, S. K. Solanki, O. C. StCyr,
and R. D. Strauss

SWx forecasting is an expression of understanding a system to the extent that its future behavior can be predicted. Two main SWx science problems are addressed in this White Paper:

(1) **Improve understanding and develop the capability to accurately forecast solar radiation hazards that affect human explorers and technology in space; and**
(2) **Improve accuracy and reliability of geomagnetic storm forecasting that protects ground-based technology and continuation of essential services.**

These two objectives can be achieved through strategic placement of SWx missions and advanced communications technology. Their strategic placement minimizes the number of total missions required and takes into account foreseeable timelines of NASA's human exploration plans, which will return humans to the Moon in the mid-2020s and send human missions to Mars in 2030 and beyond. Historically, the Earth-Sun Lagrange points (such as L1) are useful for placement of space weather missions, given their minimal maintenance requirements, and their consistent, uninterrupted long-term observations. They also provide rather benign environments as compared to missions in Earth or other planetary orbits as they avoid eclipses and exposure to radiation belts. However, space weather effects from the Sun are still present.

**Strategy for Objective 1: Securing NASA's Human Exploration of the Moon and Mars**
The sudden exposure to energetic protons from solar energetic particle (SEP) events beyond the Earth's magnetosphere and en route to the Moon or Mars can, in extreme cases, lead to acute radiation sickness, an impairing, mission-endangering condition for astronauts. Timely warnings of their impending occurrence may significantly reduce radiation exposure by allowing astronauts enough time to move to a radiation shelter. Longer-term warnings would even allow for better mission planning. Several techniques exist to provide short-term forecasts of hazards from solar energetic particle events, tailored to a variation of proton energy ranges or integral fluxes. Balch ("protons") [2008] relies upon a combination of inputs, the time-integrated and peak soft X-ray flux, along with simultaneous Types II and IV radio bursts, for determining the probability of occurrence of a proton event. Núñez ("UMASEP") [2011] exploits the rise of relativistic protons at Earth, in combination with a near-simultaneous flare X ray event from the Sun. The Air Force Research Laboratory (AFRL) uses the "PPS" model [Smart & Shea, 1979]. Needed inputs of the AFRL PPS model are solar flare locations, peak or time-integrated X-ray or radio fluxes of the flare and their times of onsets and maxima. Not currently in use due to the lack of near-real-time data is the ESPERTA concept that uses time-integrated type III radio burst intensity along with flare X-ray size and location parameters [Laurenza et al., 2009]. Moreover, StCyr et al. [2017] propose using imaging of west-limb CMEs low in the corona for SEP forecasting. In use with near-real time data since 2009 is the "REleASE" [Posner, 2007], which is based on parameters of relativistic SEP electrons for forecasting the arrival and maximum intensity of <50 MeV SEP protons at 1 AU. The REleASE system, if expanded from Earth-Sun L1 (currently ACE, soon IMAP) to **include a spacecraft stationed at Mars-Sun L1** would allow advance warnings of SEPs for the entire return trip to and from Mars [Posner & Strauss, 2020]. REleASE does not depend on remote sensing observation of the flare source but depends on an empirical forecasting matrix that currently doesn't exist for forecasts from Mars distance. The urgency for the mission to Mars-Sun L1 derives from the need to observe solar particle events from Mars distance so that the system is fully trained and working when the first human missions to Mars are undertaken. For this, the solar activity cycle needs to be taken into account.

Richardson et al. [2014] have found that about 30% of the most hazardous SEP events in the two STEREO era originate from behind the western solar limb of a 1 AU observer (including Earth). The lack of behind-the-west-limb flare observations affects the accuracy of many of the leading current SEP forecasting systems mentioned above. A combination of the methods listed would likely result in the optimum outcome. Therefore, an observatory stationed ahead of the Earth in its orbit, covering the entire **Solar Radiation Hemisphere,** as shown in Figure 1, would be needed.

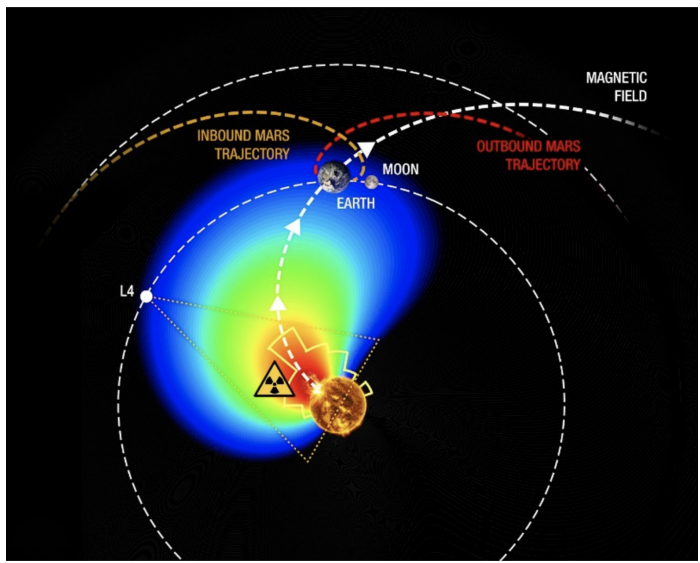

*Figure 1: The Solar Radiation Hemisphere is the relative solar hemisphere from a 1 AU observer (Earth in this case) that has the potential to severely affect its local radiation environment. It spans solar longitudes from 30E to 150W relative to the observer and is centered around and fully observable from a location 60° ahead in the observers' orbit (here the Sun-Earth L4 location). The histogram shows the relative source longitudes of all three spacecraft SEP events in the STEREO era as described in Richardson et al. [2014]. SEP event modeling is based on Strauss, Dresing & Engelbrecht [2017].*

With added flare information from Sun-Earth L4, the above short-term SEP forecasting schemes would include sufficient information to cover all SEPs that affect the Earth/Moon system, and, utilizing the Hohmann-Parker effect [Posner et al., 2013] also the outbound trajectory from Earth to Mars. Resulting short-term forecasts from Earth-Sun L1, Mars-Sun L1 and Earth-Sun L4 can save astronauts from acute radiation sickness and would limit their lifetime exposure. Depending on available shielding and radiation shelter, up to 10 cGy blood-forming organ dose can be saved in a worst-case scenario. Longer-term forecasts would depend on understanding and predicting solar activity. Recently, progress has been made in flare forecasting [e.g., Kusano et al., 2020. Wang et al., 2020, Chen et al., 2020], and it is expected that with machine learning patterns leading up to eruptions and flares can be recognized even more reliably. All these methods have in common, when applied to solar radiation forecasting, that the Solar Radiation Hemisphere is observed continuously, requiring a mission to L4.

**Strategy for Objective 2: Securing Earth from CME-Driven, Earth-Directed Geomagnetic Storms**
Earth-directed coronal mass ejections are the source of all major geomagnetic storms. A historically significant space weather event was witnessed by R. Carrington on 1 September 1859. Magnetometer records suggest a very fast, <17h [Carrington, 1859; Cliver and Svalgaard, 2004] propagation time of the CME from Sun to Earth as compared to the more typical solar wind propagation time of ~72h. Early ground-based magnetometer readings have been interpreted as equivalent to the disturbance storm time (Dst) index value of ~-850 to -1,760 [Siscoe et al., 2006; Tsurutani et al., 2003], which, although contamination by auroral currents may be possible, would so far be unsurpassed in the space age. As a context, geomagnetic storm conditions are considered severe when Dst dips below ~-150. Riley [2012] statistically analyzed occurrence rates of space weather parameters including Dst and predicted that the likelihood of occurrence during the next decade of a similar or larger storm would be ~12%. Space weather forecasting critically depends upon availability of timely and reliable observational data. Extreme space weather creates challenging conditions under which instrumentation and spacecraft may be impeded or in which parameters reach values that are outside the nominal observational range. An assessment of reliability of current and near-future space weather assets found that at least two widely spaced coronagraphs covering the Sun-Earth line, including L4, would provide reliability for Earth-bound CMEs [Posner, Hesse & StCyr, 2014]. Placement of a resilient spacecraft with in-situ instrumentation at Earth-Sun L1 would provide ground truth and a short-term forecast for such CME-driven disturbances before they encounter Earth's magnetosphere.

Moreover, it is essential to fully understand the inner heliospheric solar wind. Data-driven models such as WSA ENLIL currently depend on solar magnetic field observations from near Earth [Arge and Pizzo, 2000], which leaves large uncertainties due to the unobservable evolution of the Sun's activity behind the solar limb, which can be mitigated by placing a solar magnetograph at the Earth-Sun L5 point [Pevtsov et al., 2019]. Yet changes in activity behind the limb from Earth are not the only major source of uncertainty for inner heliosphere solar wind modeling. Changes in the intensity of the solar magnetic field near the sun's poles add uncertainty, due to the ebb and flow of their respective influence over the equatorial/ecliptic regions, where Earth is located. The Solar Orbiter mission will be the first to explore the Sun's poles through remote sensing [e.g., Solanki et al., 2019] and will add to our understanding of their behavior. In order to maximize the effectiveness of the L5-Earth-L4 set of solar remote sensing and in situ missions, we propose **launching L5 and L4 into an orbit that is inclined by ~14° with respect to the ecliptic**, so that the L4 and L5 orbits are tilted to the opposite side of the heliographic equator. This in combination with their relative longitudes will allow for observing 5/6th of the solar equator and continuous "view" of both solar poles, in effect broadening the reliable latitude coverage of magnetographic observations as shown in Figure 2. In-situ solar wind and magnetic field measurements would allow a bimonthly snapshot of the streamer and current sheet structure to be taken from three vantage points 6° apart, observations that would feed back into and advance robustness of inner heliosphere modeling. L4 and L5 will utilize optical communications for added data rate.

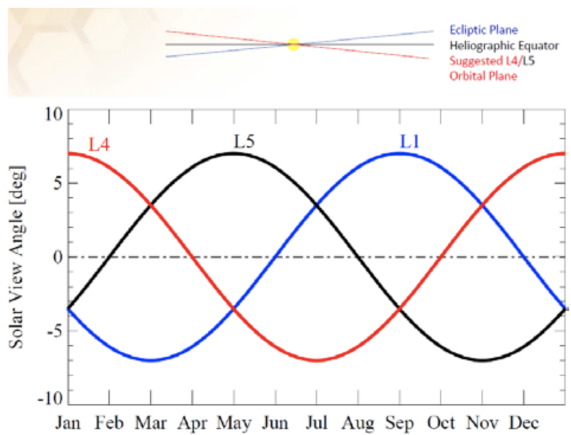

*Figure 2: Heliographic latitudes of the Earth-Sun L4, L5, and L1 missions over the course of the year. The L4 and L5 missions are injected into an inclined orbit with the same (but reversed) inclination of the ecliptic vs. the heliographic equator.*

| Objective(s) | Mission Target | Timeline | Observations supporting Objectives |
|---|---|---|---|
| **1, 2** | Earth-Sun L1 (IMAP*, SWFO*) | 2024 | Sol. Wind, Magn. Field, Energ. Particles, Radio |
| **1** | Mars-Sun L1* | 2028 | Energetic Particles |
| **1, 2** | Earth-Sun L4*, L5* | 2030 | Magnetogr., X Rays, Hsph. Imager, Radio, In Situ |
| *In Planning | *New | | |